\documentclass[12pt,a4paper,leqno]{article}

\usepackage{amssymb,exscale}
\usepackage[centertags]{amsmath}
\usepackage{amsthm}
\usepackage{graphicx}

\usepackage[dvips]{hyperref}

\numberwithin{equation}{section}

\swapnumbers
\theoremstyle{definition}
\newtheorem{theorem}[equation]{Theorem}
\newtheorem{lemma}[equation]{Lemma}

\newtheorem{corollary}[equation]{Corollary}

\renewcommand{\phi}{\varphi}

\newcommand{\I}{{\rm i}}
\newcommand{\D}{\mathrm{d}}
\newcommand{\E}{\mathrm{e}}

\renewcommand{\(}{\bigl(}
\renewcommand{\)}{\bigr)\vphantom{)}}

\newcommand{\trace}{\operatorname{trace}}
\newcommand{\Si}{\operatorname{Si}}
\newcommand{\sgn}{\operatorname{sgn}}

\newcommand{\ang}{{\text{ang}}}

\newcommand{\eps}{\varepsilon}

\newcommand{\de}{\delta}
\newcommand{\al}{\alpha}
\newcommand{\la}{\lambda}

\newcommand{\const}{{\mathrm{const}}}
\newcommand{\R}{\mathbb R}

\newcommand{\dimensional}[1]{$#1$\nobreakdash-\hspace{0pt}dimensional}

\begin{document}

\title{How often is the coordinate of a harmonic oscillator positive?} 

\author{Boris Tsirelson}

\date{}
\maketitle

\stepcounter{footnote}
\footnotetext{%
 This research was supported by \textsc{the israel science foundation}
 (grant No.~683/05).}

\begin{abstract}
The coordinate of a harmonic oscillator is measured at a time chosen at random
among three equiprobable instants: now, after one third of the period, or
after two thirds. The (total) probability that the outcome is positive depends
on the state of the oscillator. In the classical case the probability is $
\frac12 \pm \frac16 $, but in the quantum case it is $ 0.50 \pm 0.21 $.
\end{abstract}

\section*{Introduction}
The coordinate $ q(t) $ of a harmonic oscillator depends on the time $ t $,
\[
q(t) = q_0 \cos t + p_0 \sin t \, ,
\]
where $ q_0, p_0 $ are the initial coordinate and momentum; the mass is
assumed to equal $ 1 $, and the period --- to equal $ 2\pi $. In the classical
setup $ q_0 $ and $ p_0 $ are numbers, while in the quantum setup they are
operators,
\[
Q(t) = Q \cos t + P \sin t \, , \quad  [Q,P] = \I
\]
(assuming also $ \hbar = 1 $).
Here is a question trivial in the classical setup but nontrivial in the
quantum setup. We choose $ \tau $ at random from the three-element set $ \{ 0,
2\pi/3, 4\pi/3 \} $ and check, whether $ q(\tau) > 0 $ or
not. The (total, unconditional) probability of the event $ q(\tau) > 0 $
depends on the initial state of the oscillator. The question: what is the
maximum of the probability over all states? In the classical setup the
probability is either $ 1/3 $ or $ 2/3 $.
\[
\includegraphics{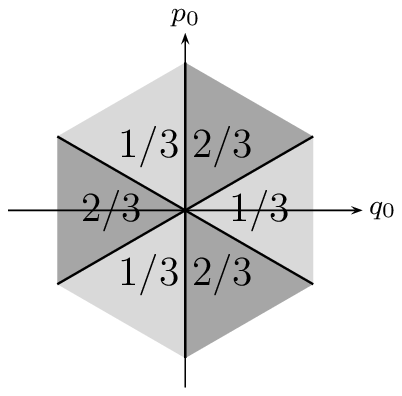}
\]
In the quantum setup the probability is
\[
\text{prob}(\psi) = \tfrac13 \langle \psi | E(0) + E(2\pi/3) +
E(4\pi/3) | \psi \rangle \, ;
\]
here $ \psi $ is the state vector, $ E(t) = \theta (Q(t)) $, and $ \theta(q) =
1 $ if $ q>0 $, otherwise $ 0 $. Thus, we want to find the upper bound of
the spectrum of the operator $ E(0) + E(2\pi/3) + E(4\pi/3) $. The question is
nontrivial, because the three terms do not commute. Note that we perform a
quantum measurement only once (at a time chosen beforehand), therefore the
impact of the measurement on the state is irrelevant.

A numeric computation reported in Sect.~\ref{sect1} shows that the upper
bound, $ \sup_\psi \text{prob}(\psi) $, of the spectrum is close to $
0.71 $. A rigorous result of Sect.~\ref{sect2} states that $ \sup_\psi
\text{prob}(\psi) < 1 $. Sect.~\ref{sect3} generalizes this result to $ E(0) +
E(s) + E(t) $.

\section[]{\raggedright Using the Wigner quasi-distribution}
\label{sect1}The Wigner function (or quasi-distribution density) $ W_\psi : \R^2 \to \R $
corresponding to a state vector $ \psi $ has several equivalent (sometimes, up
to a coefficient) definitions. The `tomographic' definition (see also
\cite{BZ}, Sect.~6.2, Th.~6.1), stipulating that $ W_\psi $ returns correct
one-dimensional distributions, is based on the equality
\begin{equation}\label{tomo}
\iint f(aq + bp) \, W_\psi(q,p) \, \D q \D p = \langle \psi | f(aQ + bP) | \psi
\rangle
\end{equation}
for all $ a,b \in \R $ and all bounded measurable functions $ f : \R \to \R
$. However, there is a catch: $ W_\psi $ need not be integrable, that is, $
\iint | W_\psi(q,p) | \, \D q \D p = \infty $ for some $ \psi $ (for example,
$ \psi(q) = (q^2+1)^{-1/3} \E^{\I q^3} $ in the Schr\"odinger
representation). Thus, even the most well-known relation $ \iint W_\psi(q,p)
\, \D q \D p = 1 $ needs a careful interpretation!

One may interprete the left-hand side of \eqref{tomo} as
\[
\lim_{\eps\to0+} \iint f(aq + bp) \, W_\psi(q,p) h((bq-ap)\eps) \, \D q \D p
\]for an appropriate weight function $ h $ such as
\[
h(x) = \E^{-x^2} \quad \text{or} \quad h(x) = \begin{cases}
 1 - |x| &\text{for $ |x| \le 1 $},\\
 0 &\text{otherwise}.
\end{cases}
\]
(It would be natural to put $ h(x) = 1 $ for $ |x| \le 1 $, otherwise $ 0
$. However, I do not know, whether this function fits or not.)
Treated this way, \eqref{tomo} holds for all $ \psi $.

Alternatively, one may characterize the whole map $ \psi \mapsto W_\psi $ by
two conditions:
\begin{itemize}
\item[(a)] there exists a dense set of state vectors $ \psi $ such that $
W_\psi $ is integrable and \eqref{tomo} is satisfied;
\item[(b)] $ W_{\psi_n} (q,p) \to W_\psi(q,p) $ (as $ n\to\infty $) uniformly
in $ q,p $ whenever $ \psi_n \to \psi $.
\end{itemize}

In the Schr\"odinger representation,
\[
W_\psi (q,p) = \frac1{2\pi} \int \psi \Big( q + \frac x 2 \Big) \overline{
\psi \Big( q - \frac x 2 \Big) } \E^{\I px} \, \D x
\]
for $ \psi \in L_2(\R) $, and of course,
\[
(Q\psi) (q) = q \psi(q) \, , \quad (P\psi) (q) = -\I \frac{\D}{\D q} \psi(q)
\]
for $ \psi $ good enough.

We have
\[
\langle \psi | E(t) | \psi \rangle = \iint \theta( q\cos t + p\sin t ) W_\psi
(q,p) \, \D q \D p = \int_{t-\pi/2}^{t+\pi/2} W_\psi^\ang (\phi) \, \D\phi \, ,
\]
where
\[
W_\psi^\ang (\phi) = \int_0^\infty W_\psi ( r\cos\phi,r\sin\phi) \, r \D r
\]
may be called the angular Wigner function. Thus,
\begin{equation}\label{1.2}
\text{prob}(\psi) = \frac13 \langle \psi | E(0) + E(2\pi/3) +
E(4\pi/3) | \psi \rangle = \frac13 \int_{-\pi}^\pi f(\phi) W_\psi^\ang (\phi)
\, \D\phi \, ,
\end{equation}
where $ f(\cdot) $ takes on two values, $1$ and $2$, as follows:
\begin{equation}\label{1.3}
\begin{gathered}\includegraphics{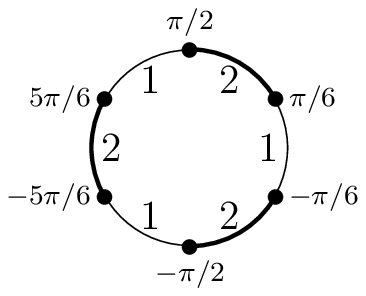}\end{gathered}
\end{equation}
The inequality $ \frac13 \le \text{prob}(\psi) \le \frac23 $ holds if $
W_\psi^\ang (\phi) \ge 0 $ for all $ \phi $, but fails in general.

Eigenvectors of the Hamiltonian are the number states $ |0\rangle, |1\rangle,
|2\rangle, \dots $;
\[
(Q^2+P^2) |n\rangle = (2n+1) |n\rangle \, .
\]
Generally,
\begin{gather*}
\psi = \sum_{n=0}^\infty c_n |n\rangle \, , \quad \sum_{n=0}^\infty |c_n|^2 =
 1 \, , \\
W_\psi (q,p) = \sum_{m,n} \bar c_m c_n w_{m,n} (q,p) \, ,
\end{gather*}
where functions $ w_{m,n} $ are defined by
\[
\iint f(aq + bp) w_{m,n} (q,p) \, \D q \D p = \langle m | f(aQ + bP) | n
\rangle \, .
\]
In fact,
\begin{equation}\label{*}
w_{m,n} (q,p) = \frac{(-1)^m}\pi \sqrt{ \frac{m!}{n!} } (2z)^{n-m}
\E^{-2|z|^2} L_m^{(n-m)} (4|z|^2)
\end{equation}
for $ m \le n $ (and $ w_{n,m} = \bar w_{m,n} $); here $ z = (q+\I p)/\sqrt2
$, and $ L_m^{(k)} $ is the associated Laguerre polynomial
\cite[Sect.~10.12]{BE},
\begin{gather*}
L_m^{(k)} (x) = (-1)^k \frac{\D^k}{\D x^k} L_{m+k}(x) \, , \\
L_n(x) = \frac1{n!} \E^x \frac{\D^n}{\D x^n} (x^n \E^{-x}) = \sum_{k=0}^n
(-1)^k \binom n k \frac{x^k}{k!} \, .
\end{gather*}
Equality \eqref{*} is a combination of \cite[(3.30)]{CG1} and
\cite[(4.36)]{CG2}.

For the angular Wigner function,
\begin{gather*}
W_\psi^\ang (\phi) = \sum_{m,n} \bar c_m c_n w_{m,n}^\ang (\phi) \, , \\
w_{m,n}^\ang (\phi) = \int_0^\infty w_{m,n} ( r\cos\phi,r\sin\phi) \, r \D r
 \, .
\end{gather*}
Using \eqref{*} we get, first,
\[
w_{m,n}^\ang (\phi) = \E^{\I(n-m)\phi} w_{m,n}^\ang (0)
\]
and second,
\begin{multline}\label{square}
w_{m,n}^\ang (0) = \int_0^\infty w_{m,n} (r,0) \, r \D r = \\
= \frac{(-1)^m}\pi \sqrt{ \frac{m!}{n!} } \int_0^\infty (\sqrt2 r)^{n-m}
 \E^{-r^2} L_m^{(n-m)} (2r^2)  \, r \D r = \\
= \frac{(-1)^{m+n}}\pi \sqrt{m!n!} \sum_{k=\max(m,n)}^{m+n} (-1)^k
2^{k-\frac{m+n}2-1} \frac{ \Gamma(k-\frac{m+n}2+1) }{ (m+n-k)!(k-m)!(k-n)! }
\end{multline}
(the latter formula holds for all $ m,n $, but the formula with $ L_m^{(n-m)}
$ holds for $ m \le n $, of course). Using the generating function
\cite[Sect.~10.12, (17)]{BE}
\[
\sum_{m=0}^\infty L_m^{(k)}(x) t^m = (1-t)^{-k-1} \exp \Big( \! -
\frac{xt}{1-t} \Big)
\]
we get also (for $ m \le n $)
\begin{multline*}
w_{m,n}^\ang (0) = \\
\frac{(-1)^m}{\pi} \sqrt{ \frac{m!}{n!} } 2^{\frac{n-m}2-1}
 \Gamma \Big( \frac{n-m}2 + 1 \Big) \frac1{m!} \frac{\D^m}{\D t^m} \bigg|_{t=0}
 \frac1{ (1-t)^{(n-m)/2} (1+t)^{(n-m)/2+1} } \, ,
\end{multline*}
which leads (by contour integration) to a formula suitable for asymptotic
analysis,
\begin{multline*}
w_{m,n}^\ang (0) = \\
\frac{(-1)^m}{\pi} \sqrt{ \frac{m!}{n!} } 2^{\frac{n-m}2-1}
 \Gamma \Big( \frac{n-m}2+1 \Big) \frac1{2\pi} \int_{-\pi}^\pi
 \frac{1-r\E^{\I\phi}}{ (1-r^2\E^{2\I\phi})^{(n-m)/2+1} } \frac{\D\phi}{
 (r\E^{\I\phi})^m } \, ;
\end{multline*}
it holds for every $ r \in (0,1) $ and is especially useful for $ r=\sqrt{m/n}
$. Here are three asymptotic results obtained this way.

First, $ \lim_{m\to\infty} w_{m,m+k}^\ang (0) = 1/(2\pi) $ for every $
k=0,1,2,\dots $

Second, let $ m,n \to \infty $, $ m/n \to c \in (0,1) $. If in addition $ m $
remains even, then $ w_{m,n}^\ang (0) \to (2\pi)^{-1} c^{-1/4} $. However, if
$ m $ remains odd, then $ w_{m,n}^\ang (0) \to (2\pi)^{-1} c^{1/4} $.

Third, let $ m $ be fixed and $ n \to \infty $, then $ w_{m,n}^\ang (0) \sim
c_m n^{1/4} $ for some $ c_m $.

Returning to \eqref{1.2} we have, first,
\begin{multline*}
\int f(\phi) W_\psi^\ang (\phi) \, \D\phi = \sum_{m,n} \bar c_m c_n \int
 f(\phi) w_{m,n}^\ang (\phi) \, \D\phi = \\
= \sum_{m,n} \bar c_m c_n w_{m,n}^\ang (0) \int f(\phi) \E^{\I(n-m)\phi} \,
\D\phi
\end{multline*}
for any $ f \in L_2(0,2\pi) $, and second,
\begin{equation}\label{squaresquare}
\int f(\phi) \E^{\I k\phi} \, \D\phi = \begin{cases}
 3\pi & \text{if } k=0 , \\
 6/k & \text{if $ (k+3)/12 $ is an integer} , \\
 -6/k & \text{if $ (k-3)/12 $ is an integer} , \\
 0 & \text{otherwise}
\end{cases}
\end{equation}
for the function $ f $ of \eqref{1.3}. The matrix elements of the
operator $ \frac13 \( E(0) + E(2\pi/3) + E(4\pi/3) \) $ in the basis of number
states are thus calculated,
\[
\frac13 \langle m | E(0) + E(2\pi/3) + E(4\pi/3) | n \rangle = \frac13
w_{m,n}^\ang (0) \int f(\phi) \E^{\I(n-m)\phi} \, \D\phi \, ,
\]
$ w_{m,n}^\ang (0) $ being given by \eqref{square} and the integral by
\eqref{squaresquare}. The spectrum of the corresponding infinite matrix
determines the possible values of $ \text{prob}(\psi) $, namely, $ \la_{\min}
\le \text{prob}(\psi) \le \la_{\max} $, where $ \la_{\min}, \la_{\max} $ are
the least and greatest elements of the spectrum.

Restricting ourselves to a finite portion $ |0\rangle, |1\rangle, \dots,
|N-1\rangle $ of the basis of number states, we get a finite matrix, $ N
\times N $, and can compute its spectrum numerically. The results follow.
\[
\begin{matrix}
 N & & 25 & 50 & 100 & 200 & 300 \\
 \la_{\max}(N) & & 0.6961 & 0.6997 & 0.7025 & 0.7045 & 0.7054
\end{matrix}
\]
Probably, the number $ \la_{\max} = \lim_N \la_{\max}(N) $ is close to $ 0.71
$. But, who knows? It could happen that $ \la_{\max} = 1 $ and, say, $
1/(1-\la_{\max}(N)) \approx 3 + 0.04 \ln N $ for large $ N $.

In Section \ref{sect2} it is shown that $ \la_{\max} \ne 1 $.

Here is the plot of the angular Wigner function $ W_\psi^\ang $, where $ \psi
= \psi_{\max}(300) $ is the \dimensional{300} eigenvector corresponding to $
\la_{\max}(300) $.
\[
\begin{gathered}\includegraphics{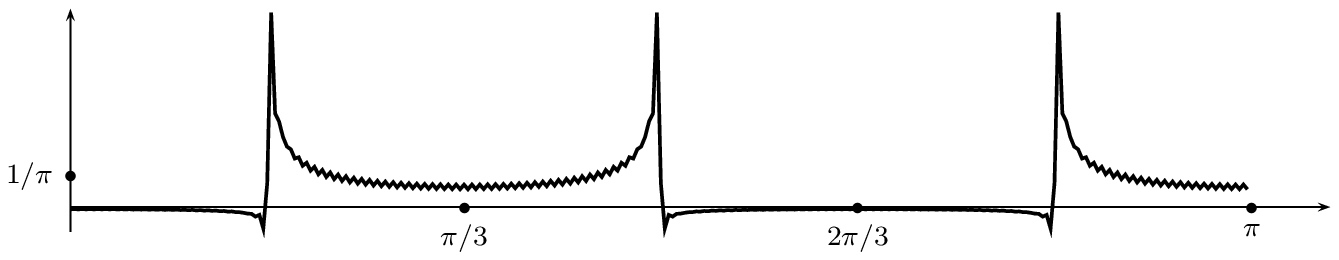}\end{gathered}
\]
Its (small) negative values are responsible for the (small) quantum violation
of the classical bound ($ 0.71 > 2/3 $).

\section[]{\raggedright Using the Weyl transform}
\label{sect2}We are interested in the spectral bounds of the operator
\[
\frac13 \( E(0) + E(2\pi/3) + E(4\pi/3) \) = \frac12 + \frac16 A \, ,
\]
where
\begin{multline*}
A = \sgn Q(0) + \sgn Q(2\pi/3) + \sgn Q(4\pi/3) = \\
= \sgn(Q) + \sgn \Big( \! -\frac12 Q + \frac{\sqrt3}2 P \Big) + \sgn \Big(
 \! -\frac12 Q - \frac{\sqrt3}2 P \Big) \, .
\end{multline*}
Note that $ (-A) $ is unitarily equivalent to $ A $ (since the symplectic
linear transformation $ (q,p) \mapsto (-q,-p) $ corresponds to a unitary
operator), therefore the spectral bounds of $ A $ are $ \pm \|A\| $, and the
spectral bounds of $ \frac12 + \frac16 A $ are
\[
\la_{\min} = \frac12 - \frac16 \|A\| \, , \quad \la_{\max} = \frac12 + \frac16
\|A\| \, .
\]
Clearly, $ \| A \| \le 3 $ (since $ \| \sgn Q(t) \| = 1 $).

\begin{theorem}\label{th}
$ \| A \| < 3 $.
\end{theorem}

In order to prove the theorem, it is sufficient to prove that the spectrum of
$ A $ does not contain $ 3 $.

The next lemma is the first step toward this goal.

\begin{lemma}\label{2.1}
The number $ 3 $ is not an eigenvalue of $ A $.
\end{lemma}

\begin{proof}
Assume the contrary, then $ 3 $ is also an eigenvalue of $ 3(\frac12+\frac16
A) $, which means
\[
\langle \psi | E(0) + E(2\pi/3) + E(4\pi/3) | \psi \rangle = 3
\]
for some state vector $ \psi $, $ \| \psi \| = 1 $. Then $ \langle \psi | E(t)
| \psi \rangle = 1 $ for $ t \in \{ 0, 2\pi/3, 4\pi/3 \} $ (since $ \langle
\psi | E(t) | \psi \rangle \le 1 $ for these $ t $).

Note that $ [Q(0),Q(t)] = [Q, Q \cos t + P \sin t] = \I\sin t $, in
particular, $ [Q(0),Q(2\pi/3)] = \frac{\sqrt3}2 \I $. Moreover, the pair of
operators $ \( \frac2{\sqrt3} Q(0), Q(2\pi/3) \) $ is unitarily equivalent to
the pair $ \( Q(0), Q(\pi/2) \) = (Q,P) $ (since every simplectic linear
transformation of the phase plane corresponds to a unitary operator). It
follows that the pair of operators $ \( E(0), E(2\pi/3) \) = \(
\theta(\frac2{\sqrt3} Q(0)), \theta(Q(2\pi/3)) \) $ is unitarily equivalent to
the pair $ \( E(0), E(\pi/2) \) = \( \theta(Q), \theta(P) \) $.

Existence of a unit vector $ \psi $ satisfying $ E(0)\psi = \psi =
E(2\pi/3)\psi $ imlies existence of a unit vector $ \psi_1 $ satisfying $
\theta(Q)\psi_1 = \psi_1 = \theta(P)\psi_1 $. In other words, we get a
wavefunction $ \psi_1 \in L_2(\R) $ concentrated on the halfline $ (0,\infty)
$, whose Fourier transform is also concentrated on $ (0,\infty) $. However,
this is forbidden by a well-known theorem of F.\ and M.\ Riesz \cite[Part One,
Chapter 1, \S 1]{HJ}.
\end{proof}

In spite of Lemma \ref{2.1}, the number $ 3 $ could belong to the spectrum of
$ A $. For example, consider the operator $ B = \( E(0) + E(\pi/2) \) / 2 = \(
\theta(Q) + \theta(P) \) / 2 $. The number $ 1 $ is not an eigenvalue of $ B $
(recall the proof of Lemma \ref{2.1}), but still belongs to the spectrum of $
B $, since $ \langle \psi_n | B | \psi_n \rangle \to 1 $ for coherent states $
\psi_n $ such that $ \langle \psi_n | Q | \psi_n \rangle \to +\infty $, $
\langle \psi_n | P | \psi_n \rangle \to +\infty $.

In contrast, the spectrum of $ A $ is discrete, except for two accumulation
points, $ -1 $ and $ 1 $. In order to prove this claim we prove that the
operator $ A^2-1 $ is compact, moreover, belongs to the Hilbert-Schmidt class,
\[
\trace \( (A^2-1)^2 \) < \infty \, .
\]
We do it by calculating the Weyl transform of $ A^2 $.

In general, the Weyl transform (or Weyl symbol) of a bounded operator $ B : H
\to H $ is defined as a Schwartz distribution $ f $ on $ \R^2 $ such that the
equality
\[
\iint f(q,p) W_\psi(q,p) \, \D q \D p = \langle \psi | B | \psi \rangle
\]
holds for all $ \psi $ of a dense set of state vectors. It is assumed
that for each $ \psi $ of this set, the Wigner function $ W_\psi $ is a
rapidly decreasing, infinitely differentiable function. It is well-known that
$ f $ is uniquely determined by $ B $.

However, we do not need Schwartz distributions; we restrict ourselves to
bounded measurable functions $ f : \R^2 \to \R $ (and the corresponding
operators $ B $). Accordingly, we do not need the differentiability of $
W_\psi $. Also, we waive the rapid decrease of $ W_\psi $, demanding only
$ \iint |W_\psi(q,p)| \, \D q \D p < \infty $.

Clearly, the function $ (q,p) \mapsto f(aq+bp) $ is the Weyl transform of the
operator $ f(aQ+bP) $, for all $ a,b \in \R $ and all bounded measurable
functions $ f : \R \to \R $.

It is well-known (see \cite{CG1}, (3.14)) that
\begin{equation}\label{wk}
\frac1{2\pi} \iint |f(q,p)|^2 \, \D q \D p = \trace (B^* B) \, ,
\end{equation}
thus, an operator belongs to the Hilbert-Schmidt class if and only if its
Weyl transform belongs to $ L_2(\R^2) $.

\begin{lemma}\label{2.2}
The function $ (q,p) \mapsto (2/\pi) \Si(2qp) $ is the Weyl transform of the
operator $ (\sgn Q) \circ (\sgn P) $.
\end{lemma}

Here $ \Si(x) = \int_0^x \frac{\sin u}u \, \D u $, and $ A \circ B = (AB+BA)/2
$.

The (rigorous) proof grows from a non-rigorous argument shown below before the
proof. We have
\begin{gather*}
\sgn Q = -\frac\I\pi \int_{-\infty}^\infty \E^{\I tQ} \, \frac{\D t}t \quad
 \text{(and the same for $ P $)} \, ; \\
\E^{\I tQ} \circ \E^{\I sP} = \Big( \cos \frac{st}2 \Big) \E^{\I tQ + \I sP}
 \, ; \\
(\sgn Q) \circ (\sgn P) = -\frac1{\pi^2} \iint \frac{\D s}s \frac{\D t}t \Big(
 \cos \frac{st}2 \Big) \E^{\I(tQ+sP)} \, .
\end{gather*}
The Weyl transform of the operator $ \E^{\I(tQ+sP)} $ is the function $ (q,p)
\mapsto \E^{\I(tq+sp)} $. Thus, the Weyl transform of the operator $ (\sgn
Q) \circ (\sgn P) $ is the function
\begin{multline*}
f(q,p) = -\frac1{\pi^2} \iint \frac{\D s}s \frac{\D t}t \Big( \cos \frac{st}2
 \Big) \E^{\I(tq+sp)} = \\
= \frac4{\pi^2} \iint_{s>0,t>0} \frac{\D s}s \frac{\D t}t \Big( \cos
 \frac{st}2 \Big) \sin tq \sin sp = \\
= \frac4{\pi^2} \int_0^\infty \frac{\D s}s \sin sp \int_0^\infty \frac{\D t}t
 \cos \frac s 2 t \sin qt = \\
= \frac4{\pi^2} \int_0^\infty \frac{\D s}s \sin sp \int_0^\infty
 \frac{\D t}t \bigg( \frac12 \sin \Big( q+\frac s 2 \Big) t + \frac12 \sin
 \Big( q-\frac s 2 \Big) t \bigg) = \\
= \frac4{\pi^2} \int_0^\infty \frac{\D s}s \sin ps \cdot \bigg( \frac\pi4 \sgn
 \Big( q+\frac s 2 \Big) + \frac\pi4 \sgn \Big( q-\frac s 2 \Big) \bigg) = \\
= \frac4{\pi^2} \cdot \frac\pi2 \cdot \sgn q \cdot \int_0^{2|q|} \frac{\D s}s
 \sin ps = \frac2\pi \Si(2qp) \, .
\end{multline*}

In order to convert the calculation above into a proof we need the equality
\begin{multline*}
-\frac1{\pi^2} \iint \frac{\D s}s \frac{\D t}t \Big( \cos \frac{st}2
 \Big) \iint \D q \D p \, \E^{\I(tq+sp)} W_\psi (q,p) = \\
= -\frac1{\pi^2} \iint \D q \D p \, W_\psi (q,p) \iint \frac{\D s}s \frac{\D
 t}t \Big( \cos \frac{st}2 \Big) \E^{\I(tq+sp)} \, .
\end{multline*}
Unfortunately, it does not follow from Fubini's theorem, since for large $ s,t
$ the integrand decays too slowly. (Also small $ s,t $ make a trouble, but
less serious.) A cutoff is used below.

\begin{proof}[Proof of Lemma \textup{\ref{2.2}}]
We have for all $ q $
\[
-\frac\I\pi \int_{\frac1n<|t|<n} \E^{\I tq} \, \frac{\D t}t = \frac2\pi
 \int_{1/n}^n \frac{\sin qt}t \, \D t = \frac2\pi \( \Si(qn) - \Si(q/n) \)
 \to \sgn q
\]
(as $ n \to \infty $), and these functions are bounded in $ q $, uniformly in
$ n $. The same holds for $ P $, and we get
\begin{multline*}
\langle \psi | (\sgn P) (\sgn Q) | \psi \rangle = \\
= \Big( \! -\frac\I\pi \Big)^2 \lim_{m\to\infty} \lim_{n\to\infty}
 \bigg\langle \psi \bigg| \bigg( \int_{\frac1m<|s|<m} \frac{\D s}s \E^{\I sP}
 \bigg) \bigg( \int_{\frac1n<|t|<n} \frac{\D t}t \E^{\I tQ} \bigg) \bigg| \psi
 \bigg\rangle \, .
\end{multline*}
Therefore
\begin{multline*}
\langle \psi | (\sgn P) \circ (\sgn Q) | \psi \rangle = \\
= -\frac1{\pi^2} \lim_{m\to\infty} \lim_{n\to\infty} \int_{\frac1m<|s|<m}
 \frac{\D s}s \int_{\frac1n<|t|<n} \frac{\D t}t \Big( \cos \frac{st}2 \Big)
 \langle \psi | \E^{\I(tQ+sP)} | \psi \rangle = \\
= -\frac1{\pi^2} \lim_{m\to\infty} \lim_{n\to\infty} \int_{\frac1m<|s|<m}
 \frac{\D s}s \int_{\frac1n<|t|<n} \frac{\D t}t \Big( \cos \frac{st}2 \Big)
 \iint \D q \D p W_\psi (q,p) \E^{\I(tq+sp)} = \\
= -\frac1{\pi^2} \lim_{m\to\infty} \lim_{n\to\infty} \iint \D q \D p W_\psi
 (q,p) \int_{\frac1m<|s|<m} \frac{\D s}s \int_{\frac1n<|t|<n} \frac{\D t}t
 \Big( \cos \frac{st}2 \Big) \E^{\I tq} \E^{\I sp} = \\
= -\frac1{\pi^2} (2\I)^2 \lim_{m\to\infty} \lim_{n\to\infty} \iint \D q \D p
W_\psi (q,p) \int_{1/m}^m \frac{\D s}s \int_{1/n}^n \frac{\D t}t \Big(
\cos \frac{st}2 \Big) \sin qt \sin ps \, .
\end{multline*}
However,
\begin{multline*}
\int_{1/n}^n \frac{\D t}t \Big( \cos \frac{st}2 \Big) \sin qt = \\
= \frac12 \Si \bigg( \Big( q + \frac s 2 \Big) n \bigg) - \frac12 \Si \bigg(
 \Big( q + \frac s 2 \Big) \frac1n \bigg) + \frac12 \Si \bigg( \Big( q - \frac
 s 2 \Big) n \bigg) - \frac12 \Si \bigg( \Big( q - \frac s 2 \Big) \frac1n
 \bigg) \to \\
\to \frac\pi4 \sgn \Big( q + \frac s 2 \Big) + \frac\pi4 \sgn \Big( q - \frac
s 2 \Big) \quad \text{as } n \to \infty
\end{multline*}
for all $ s,q $, and these functions are bounded in $ s,q $, uniformly in $ n
$. Taking into account that
\[
\iint \D q \D p \, | W_\psi (q,p) | \int_{1/m}^m \frac{\D s}s |\sin ps| <
\infty
\]
we get
\begin{multline*}
\langle \psi | (\sgn P) \circ (\sgn Q) | \psi \rangle = \\
= \frac4{\pi^2} \lim_{m\to\infty} \iint \D q \D p W_\psi (q,p) \int_{1/m}^m
 \frac{\D s}s \sin ps \cdot \bigg( \frac\pi4 \sgn \Big( q + \frac s 2 \Big) +
 \frac\pi4 \sgn \Big( q - \frac s 2 \Big) \bigg) \, .
\end{multline*}
If $ 2|q| > \frac1m $ then
\begin{multline*}
\int_{1/m}^m \frac{\D s}s \sin ps \cdot \bigg( \frac\pi4 \sgn \Big( q + \frac
 s 2 \Big) + \frac\pi4 \sgn \Big( q - \frac s 2 \Big) \bigg) = \\
= \frac\pi2 (\sgn q) \int_{1/m}^{\min(m,2|q|)} \frac{\D s}s \sin ps = \\
= \frac\pi2 (\sgn q) (\sgn p) \( \Si(|p| \min(m,2|q|)) - \Si(|p|/m) \) \, .
\end{multline*}
Otherwise, if $ 2|q| \le \frac1m $, the integral vanishes. We see that
\[
\int_{1/m}^m \frac{\D s}s \sin ps \cdot \bigg( \frac\pi4 \sgn \Big( q + \frac
s 2 \Big) + \frac\pi4 \sgn \Big( q - \frac s 2 \Big) \bigg) \to \frac\pi2
\Si(2qp) \quad \text{as } m \to \infty
\]
for all $ q,p $, and these functions are bounded in $ q,p $, uniformly in $ m
$. Taking into account that $ \iint \D q \D p \, | W_\psi (q,p) | < \infty $
we get
\[
\langle \psi | (\sgn P) \circ (\sgn Q) | \psi \rangle = \frac2\pi \iint \D q
\D p \, W_\psi (q,p) \Si(2qp) \, .
\]
\end{proof}

We return to the operators $ Q(t) = Q \cos t + P \sin t $ and recall that the
pair of operators $ \( Q(s), \frac1{\sin(t-s)} Q(t) \) $ is unitarily
equivalent to the pair $ (Q,P) $ whenever $ \sin(t-s) \ne 0 $ (as was noted in
the proof of Lemma \ref{2.1} for a special case). Thus, Lemma \ref{2.2}
implies the following.

\begin{corollary}
\begin{sloppypar}
The function
\[
(q,p) \mapsto \frac2\pi \Si \Big( \frac2{|\sin(t-s)|} (q\cos s+p\sin s) (q\cos
t+p\sin t) \Big)
\]
is the Weyl transform of the operator $ \( \sgn Q(s) \) \circ \( \sgn Q(t) \)
$, whenever $ {\sin(t-s)} \ne 0 $.
\end{sloppypar}
\end{corollary}

\begin{corollary}\label{2.4}
The function
\begin{multline*}
f(q,p) = \frac2\pi \Si \bigg( \frac4{\sqrt3} q \cdot \frac{-q+p\sqrt3}2 \bigg)
 + \frac2\pi \Si \bigg( \frac4{\sqrt3} q \cdot \frac{-q-p\sqrt3}2 \bigg) + \\
+ \frac2\pi \Si \bigg( \frac4{\sqrt3} \cdot \frac{-q+p\sqrt3}2 \cdot
 \frac{-q-p\sqrt3}2 \bigg)
\end{multline*}
is the Weyl transform of the operator
\begin{multline*}
\frac12 ( A^2-3 ) = \( \sgn Q(0) \) \circ \( \sgn Q(2\pi/3) \) + \\
+ \( \sgn Q(0) \) \circ \( \sgn Q(4\pi/3) \) + \( \sgn Q(2\pi/3) \) \circ \(
 \sgn Q(4\pi/3) \) \, .
\end{multline*}
\end{corollary}

\begin{lemma}\label{2.5}
\[
\iint \( f(q,p)+1 \)^2 \, \D q \D p < \infty \, .
\]
\end{lemma}

\begin{proof}
Taking into account that $ f $ is invariant under the rotation by $ \pi/3 $
and the reflection $ (q,p) \mapsto (-q,p) $ we may integrate only over the
domain $ 0 < q\sqrt3 < p $. We divide the domain into a bounded domain (whose
contribution is evidently finite) and two unbounded domains, one being $ p > 1
$, $ 0 < q < p^{-1/3} $, the other $ p > 3 $, $ p^{-1/3} < q < p/\sqrt3 $.

Using the inequality
\[
\Big| \Si(x) - \frac\pi2 \Big| \le \frac2x \quad \text{for } 0<x<\infty
\]
(it holds, since $ \int_x^\infty \frac{\sin u}u \, \D u = \frac{\cos x}x -
\int_x^\infty \frac{\cos u}{u^2} \, \D u $), we get for $ 0 < q < p/\sqrt3 $
\begin{multline*}
|f(q,p)+1| \le \bigg| \frac2\pi \Si \Big( \frac4{\sqrt3} q \cdot
 \frac{p\sqrt3-q}2 \Big) - 1 \bigg| + \bigg| 1 - \frac2\pi \Si \Big(
 \frac4{\sqrt3} q \cdot \frac{p\sqrt3+q}2 \Big) \bigg| + \\
+ \bigg| 1 - \frac2\pi \Si \Big( \frac4{\sqrt3} \cdot
 \frac{p\sqrt3-q}2 \cdot \frac{p\sqrt3+q}2 \Big) \bigg| \le \\
\le \frac4\pi \cdot \frac{\sqrt3}4 \cdot \frac2{q(p\sqrt3-q)} + \frac4\pi
 \cdot \frac{\sqrt3}4 \cdot \frac2{q(p\sqrt3+q)} + \frac4\pi \cdot
 \frac{\sqrt3}4 \cdot \frac4{(p\sqrt3-q)(p\sqrt3+q)} \le \\
\le \frac{\sqrt3}\pi \Big( \frac2{qp\sqrt3} + \frac2{qp\sqrt3\cdot2/3} +
 \frac4{p\sqrt3(2/3)\cdot3q} \Big) \le \frac{\const}{pq} \, ,
\end{multline*}
thus
\[
\int_3^\infty \D p \int_{p^{-1/3}}^{p/\sqrt3} \D q \( f(q,p)+1 \)^2 \le \const
\cdot \int_3^\infty \D p \int_{p^{-1/3}}^\infty \D q \frac1{p^2 q^2} < \infty
\, .
\]

Using also the inequality
\[
| \Si(x) - \Si(y) | \le |x-y| \quad \text{for } x,y \in \R
\]
(it holds, since $ | \frac{\sin u}u | \le 1 $), we get for $ 0 < q < p/\sqrt3
$
\begin{multline*}
|f(q,p)+1| \le \bigg| \frac2\pi \Si \Big( \frac4{\sqrt3} q \cdot
 \frac{p\sqrt3-q}2 \Big) - \frac2\pi \Si \Big( \frac4{\sqrt3} q \cdot
 \frac{p\sqrt3+q}2 \Big) \bigg| + \\
+ \bigg| 1 - \frac2\pi \Si \Big( \frac4{\sqrt3} \cdot \frac{p\sqrt3-q}2 \cdot
 \frac{p\sqrt3+q}2 \Big) \bigg| \le \\
\le \frac2\pi \cdot \frac4{\sqrt3} q^2 + \frac4\pi \cdot \frac{\sqrt3}4 \cdot
 \frac4{(p\sqrt3-q)(p\sqrt3+q)} \le \const \cdot \Big( q^2 + \frac1{p^2} \Big)
 \, ,
\end{multline*}
thus
\[
\int_1^\infty \D p \int_0^{p^{-1/3}} \D q \( f(q,p)+1 \)^2 \le \const
\cdot \int_1^\infty \D p \int_0^{p^{-1/3}} \D q \Big( q^4 + \frac1{p^4} \Big)
< \infty \, .
\]
\end{proof}

\begin{proof}[Proof of Theorem \textup{\ref{th}}]
By \eqref{wk}, Corollary \ref{2.4} and Lemma \ref{2.5},
\[
\trace (A^2-1) = \frac1{2\pi} \iint | 2 (f(q,p)+1) |^2 \, \D q \D p < \infty
\, ,
\]
therefore every spectral point of $ A $ different from $ \pm1 $ is an
eigenvalue. By Lemma \ref{2.1}, the number $ 3 $ does not belong to the
spectrum of $ A $. It remains to use the remark after the theorem.
\end{proof}

\section[]{\raggedright Some generalizations}
\label{sect3}\begin{lemma}\label{3.1}
Let $ s,t \in \R $ satisfy $ s < \pi < t < s+\pi $. Then
\[
\| \sgn Q(0) + \sgn Q(s) + \sgn Q(t) \| < 3 \, .
\]
\end{lemma}

\begin{proof}
Similarly to Sect.~\ref{sect2} we introduce the operator
\[
A_{s,t} = \sgn Q(0) + \sgn Q(s) + \sgn Q(t) \, .
\]
Similarly to Lemma \ref{2.1}, the number $ 3 $ is not an eigenvalue of $ A
$. Thus, it is sufficient to prove that $ \trace \( (A^2-1)^2 \) < \infty
$. Similarly to Corollary \ref{2.4} we calculate the Weyl transform $ f_{s,t}
$ of $ \frac12 (A_{s,t}^2-3) $. It remains to prove that $ \iint \(
f_{s,t}(q,p)+1 \)^2 \, \D q \D p < \infty $.

We observe that
\[
s - \frac\pi2 < \frac\pi2 < t - \frac\pi2 < s + \frac\pi2 < \frac{3\pi}2 < t +
\frac\pi2 < 2\pi + s - \frac\pi2 \, ,
\]
choose $ \al_1,\dots,\al_6 $ such that
\begin{multline*}
\al_1 < s - \frac\pi2 < \al_2 < \frac\pi2 < \al_3 < t - \frac\pi2 < \al_4 < s
 + \frac\pi2 < \\
< \al_5 < \frac{3\pi}2 < \al_6 < t + \frac\pi2 < \al_1 + 2\pi
\end{multline*}
and divide the plane into six sectors by the six rays $ (q,p) =
(r\cos\al_k,r\sin\al_k) $, $ r>0 $, $ k=1,\dots,6 $. The three lines $ q=0 $,
$ q\cos s + p\sin s = 0 $, $ q\cos t + p\sin t = 0 $ give us six more rays,
and each of these rays is contained in one (and only one) of the six
sectors. We estimate the integral over a neighborhood of the ray (within the
sector) similarly to $ \int \D p \int_0^{p^{-1/3}} \D q (\dots) $ in the proof
of Lemma \ref{2.5}, and the integral over the rest of the sector similarly to
$ \int \D p \int_{p^{-1/3}}^{p/\sqrt3} \D q (\dots) $ in the proof of Lemma
\ref{2.5}.
\end{proof}

\begin{theorem}
For every $ \eps > 0 $ there exists $ \de > 0 $ such that
\[
\| \sgn Q(0) + \sgn Q(s) + \sgn Q(t) \| \le 3-\de
\]
for all $ s,t $ satisfying $ s \le \pi-\eps \le t-2\eps \le s-3\eps+\pi $.
\end{theorem}

\begin{proof}
The estimates needed in the proof of Lemma \ref{3.1} are uniform over all such
pairs $ (s,t) $. Thus,
\[
\iint_{q^2+p^2>r^2} \( f_{s,t}(q,p)+1 \)^2 \, \D q \D p \to 0 \quad \text{as }
r \to \infty
\]
uniformly in these $ s,t $. Also, $ f_{s,t}(q,p) $ is continuous in $ s,t $
for any fixed $ q,p $. Therefore $ f_{s,t} $ treated as an element of $
L_2(\R^2) $ depends continuously on $ s,t $. Taking into account the isometric
correspondence between Hilbert-Schmidt operators and their Weyl symbols we see
that the Hilbert-Schmidt operator $ A_{r,s}^2-1 $ depends continuously on $
s,t $. It follows that $ \| A_{s,t}^2-1 \| $ is continuous in $ s,t $ (the
usual operator norm is meant, not the Hilbert-Schmidt norm). By compactness,
the norm reaches its maximum at some point $ (s_0,t_0) $ of the considered set
of pairs $ (s,t) $. By Lemma \ref{3.1}, $ \| A_{s_0,t_0} \| < 3 $. Therefore $
\| A_{s_0,t_0}^2-1 \| < 8 $. We take $ \de>0 $ such that $ (3-\de)^2-1 \ge \|
A_{s_0,t_0}^2-1 \| $ and $ (3-\de)^2-1 > 1 $. Then $ \| A_{s,t}^2-1 \| \le
(3-\de)^2-1 $ implies $ \| A_{s,t} \| \le 3-\de $.
\end{proof}

\bigskip
\filbreak
{
\small
\begin{sc}
\parindent=0pt\baselineskip=12pt
\parbox{4in}{
Boris Tsirelson\\
School of Mathematics\\
Tel Aviv University\\
Tel Aviv 69978, Israel
\smallskip
\par\quad\href{mailto:tsirel@post.tau.ac.il}{\tt
 mailto:tsirel@post.tau.ac.il}
\par\quad\href{http://www.tau.ac.il/~tsirel/}{\tt
 http://www.tau.ac.il/\textasciitilde tsirel/}
}

\end{sc}
}
\filbreak

\end{document}